\documentclass{ifacconf}

\usepackage{graphicx} 
\usepackage[cmex10]{amsmath} 
\usepackage{amssymb}  
\usepackage{natbib}        
\usepackage{algorithm}
\usepackage{algpseudocode}
\usepackage{microtype,url}  
\usepackage{BOONDOX-ds}
\usepackage{pgfplots} 
\usepgfplotslibrary{groupplots} 
\usepgfplotslibrary{fillbetween}
\usepackage{tikz} 
\usetikzlibrary{patterns,calc,decorations.pathmorphing,decorations.markings, arrows, arrows.meta} 

\renewcommand{\theenumi}{(\roman{enumi})}

\renewcommand*\descriptionlabel[1]{\hspace\labelsep\normalfont#1}

\newtheorem{theorem}{Theorem}

\newtheorem{corollary}{Corollary}

\newtheorem{remark}{Remark}

\newtheorem{definition}{Definition}

\newcommand {\nn}{\nonumber}
\newcommand{\beq}{\begin{equation}}
\newcommand{\eeq}{\end{equation}}
\newcommand {\bseq}{\begin{subequations}}
\newcommand {\eseq}{\end{subequations}}
\newcommand {\bma}{\left[}
\newcommand {\ema}{\right]}

\newcommand {\R}{\mathbb{R}} 	    
\newcommand {\Rge}{\mathbb{R}_{+}} 	
 	
\newcommand {\Co}{\mathbb{C}} 	   
\renewcommand{\L}{\mathcal{L}}		
	
\newcommand{\LTp}{\mathcal{L}_{2,p}}	
	
\renewcommand {\H}{\mathcal{H}}

\renewcommand {\S}{\mathbb{S}}

\renewcommand{\Re}{\mathbf{Re}}

\newcommand{\transpose}{\mathsf{T}}

\newcommand{\norm}[1]{\lVert #1\rVert}

\newcommand{\diag}{\operatorname{diag}}

\newcommand{\spectrum}[1]{{\sigma({#1})}}
\newcommand{\overbar}[1]{\mkern 1.5mu\overline{\mkern-1.5mu#1\mkern-1.5mu}\mkern 1.5mu}

\newcommand{\ine}{\operatorname{In}}

\DeclareMathOperator{\sat}{sat}

\newcommand{\TAC}{\textit{{IEEE} Trans. Autom.
  Control}}\newcommand{\SCL}{\textit{Syst. Control
  Lett.}}\newcommand{\IJC}{\textit{Int. J.
  Control}}\newcommand{\ACC}{\textit{Proc. Amer. Control
  Conf.}}\newcommand{\CDC}[1]{\textit{Proc. {#1} Conf. Decision
  Control}}\newcommand{\NOLCOS}[1]{\textit{Proc. {#1} IFAC Symp. Nonlinear
  Control Syst.}}

\begin{document}

\begin{frontmatter}

\title{{Model reduction by balanced~truncation\\ of~dominant~Lure~systems}\thanksref{footnoteinfo}} 

\thanks[footnoteinfo]{\noindent The research leading to these results has received funding from the European Research Council under the Advanced ERC Grant Agreement Switchlet n. 670645  and from the Royal Society Research Grant RGS$\backslash$R1$\backslash$191308.}

\author{Alberto Padoan, Fulvio Forni, Rodolphe Sepulchre} 

\address{Department of Engineering, University of Cambridge,\\ 
							Cambridge, CB2 1PZ,  UK \\ 
							Email: {\tt\{a.padoan|f.forni|r.sepulchre\}@eng.cam.ac.uk}}

\begin{abstract}  
\noindent 
The paper presents a model reduction framework geared towards the analysis and design of systems that switch and oscillate. While such phenomena are ubiquitous in nature and engineering, model reduction methods are not well developed for non-equilibrium behaviors. The proposed framework addresses this need by exploiting recent advances on dominance theory. Classical balanced truncation for linear time-invariant systems is used to develop a dominance-preserving model reduction method for Lure systems, \textit{i.e.}\! systems that can be decomposed as the feedback interconnection of a linear system and a static nonlinearity. The  method   is illustrated by  approximating the oscillatory behavior of a  discretized heat flow control system.
\end{abstract}
\end{frontmatter}

\section{Introduction}

The growing importance of simulation tools in analysis and design has directed
the interest of the research community towards building accurate,  yet 
inexpensive,  mathematical models.
The need of  accuracy  leads  invariably to the incorporation of a large number of state variables. This, in turn, leads to large scale models which in general may be difficult to simulate owing to time or storage constraints. Model reduction methods alleviate this issue by constructing simplified models which retain prescribed features of the original system~\citep{antoulas2005approximation}.

Model reduction methods for linear systems belong broadly to two main classes~\citep{antoulas2005approximation}. The first class makes use of the  singular value decomposition~\citep{moore1981principal,glover1984all},  while the second class is based on the concept of moment matching or on Krylov projectors~\citep{georgiou1983partial,antoulas1990solution,georgiou1999interpolation}. While moment matching methods are generally more efficient and reliable from a numerical point of view, methods based on the singular value decomposition not only preserve important features of the original system, but also offer error bounds~\citep{antoulas2005approximation}.   

Over the past decades, several nonlinear counterparts of linear model reduction methods have been developed~\citep{berkooz1993proper,scherpen1993balancing,hahn2002improved,astolfi2010model}. 
This line of research has mostly focused on the construction of reduced order models for \textit{local} behaviors around equilibria. However, the problem of approximating the \textit{global} behavior of a nonlinear system is still open. Besselink and co-authors have addressed this problem in~\citep{besselink2009error,besselink2013model} and shown that (incremental) stability properties can be preserved for systems that can be decomposed as the feedback interconnection of a linear system and a nonlinear system. The  main idea is to reduce the linear dynamics using standard model reduction methods and to impose (incremental) small gain conditions to guarantee existence, uniqueness, and incremental stability   of a  steady-state equilibrium solution.

The present work extends the approach of~\citep{besselink2009error} to multistable and oscillatory Lure systems 
using dominance theory~\citep{forni2018differential,felix2018analysis,padoan2019feedback,padoan2019norm}.
The goal is to develop a model reduction theory for non-equilibrium behaviors.
 The proposed framework   is based on two key ingredients: 
(i) the dynamics is split into dominant and non-dominant components, and standard model reduction methods are used to reduce the non-dominant one; 
(ii) the asymptotic behavior of the closed-loop system is characterized using small gain conditions from dominance theory. These conditions guarantee that the behavior of the original system is well captured by that of the reduced order model when the approximation error is small. 
The framework is illustrated by developing a balanced truncation method for dominant Lure systems inspired by the method in~\citep{besselink2009error}.

The remainder of the paper is organized as follows. 
Section~\ref{sec:problem-formulation} provides the problem formulation.
Section~\ref{sec:preliminaries} recalls  some   preliminary results  from   dominance theory. 
Section~\ref{sec:main-results} contains the main results of the paper, where  a general model reduction framework for dominant Lure systems is presented. 
Section~\ref{sec:mr}  discusses a balanced truncation method for dominant Lure systems inspired by the method presented in~\citep{besselink2009error}.   
Section~\ref{sec:example} provides an illustrative example,   in which
the oscillatory behavior of a discretized heat flow control system
needs to be approximated.   
Section~\ref {sec:conclusion} summarizes our results  and outlines future research directions.

\section{Problem formulation} \label{sec:problem-formulation}

Consider a continuous-time, single-input, single-output, time-invariant  \textit{Lure system} described by the equations 
\beq \label{eq:system-Lure-mr}
\dot{x} = Ax +B_u u +B_w w, \, y = C_y x,  \, z = C_z x,  \, w = - \varphi(z),  \! \!
\eeq
in which ${x\in\R^n}$, ${u\in\R}$, ${y\in\R}$,  ${w\in\R}$, ${z\in\R}$, ${A\in\R^{n \times n}}$, ${B_u\in\R^{n \times 1}}$, ${B_w\in\R^{n \times 1}}$, ${C_y\in\R^{1\times n}}$, and ${C_z\in\R^{1\times n}}$ are constant matrices, and ${\varphi:\R \to \R}$ is a continuously differentiable function\footnote{Local Lipschitz continuity would be sufficient. Continuous differentiability is assumed to simplify the exposition.} such that ${\varphi(0) =0}$. Let 
\beq \nn
G(s) 
=
C (sI-A)^{-1} B
= 
\bma
\begin{array}{cc}
G_{yu}(s) & G_{yw}(s)\\
G_{zu}(s) & G_{zw}(s)
\end{array}
\ema
,
\eeq
in which ${B = [\, B_{u} \ B_{w} \,]}$ and  ${C = [\, C_{y}^{\transpose} \ C_{z}^{\transpose} \,]^{\transpose}}$, respectively.

\begin{figure}[h!]
\centering
\begin{tikzpicture}[scale=.75, black,every node/.style={transform shape}]
 
\draw[line width = .5 pt]  (10.3125,3.5) rectangle (11.3125,2.5);
\node at (10.8125,3) {$G$};
\draw[rounded corners,line width = .5 pt]  (10.3125,2.3125) rectangle (11.3125,1.3125);
\node at (10.8125,1.8125) {$-\varphi(\cdot)$};
\draw[-latex, line width = .5 pt] (11.3125,2.75) -- (12.3125,2.75) -- (12.3125,1.8125) -- (11.3125,1.8125);
\draw[-latex, line width = .5 pt] (10.3125,1.8125) -- (9.3125,1.8125) -- (9.3125,2.75) -- (10.3125,2.75) ;
\draw [-latex, line width = .5 pt](9.3125,3.25) -- (10.3125,3.25);
\node at (9.8125,3.5) {$u$};
\node at (11.8125,3.5) {$y$};
\node at (9.8125,2.5) {$w$};
\node at (11.8125,2.5) {$z$}; 
\draw [-latex,line width = .5 pt](11.3125,3.25) -- (12.3125,3.25);

\draw[line width = .5 pt]  (16.6875,3.5) rectangle (17.6875,2.5);
\node at (17.1875,3) {$\hat{G}$};
\draw[rounded corners,line width = .5 pt]  (16.6875,2.3125) rectangle (17.6875,1.3125);
\node at (17.1875,1.8125) {$-\varphi(\cdot)$};
\draw[-latex, line width = .5 pt] (17.6875,2.75) -- (18.6875,2.75) -- (18.6875,1.8125) -- (17.6875,1.8125);
\draw[-latex, line width = .5 pt] (16.6875,1.8125) -- (15.6875,1.8125) -- (15.6875,2.75) -- (16.6875,2.75) ;
\draw [-latex, line width = .5 pt](15.6875,3.25) -- (16.6875,3.25);
\node at (16.1875,3.5) {$\hat{u}$};
\node at (18.1875,3.5) {$\hat{y}$};
\node at (16.1875,2.5) {$\hat{w}$};
\node at (18.1875,2.5) {$\hat{z}$}; 
\draw [-latex,line width = .5 pt](17.6875,3.25) -- (18.6875,3.25);

\draw [thick, -latex](12.8125,2.5) -- (15.1875,2.5);
\node at (14,3) { Model reduction};
 
\end{tikzpicture} 
\centering
\caption{Diagrammatic illustration of the original system~\eqref{eq:system-Lure-mr} (left) and of the reduced order model~\eqref{eq:system-ROM} (right).
}
\label{fig:model_reduction}
\end{figure}
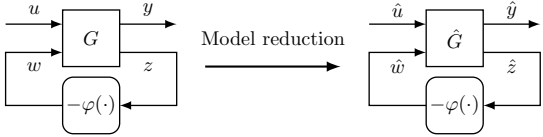

Suppose we wish to construct a reduced order model   of   order ${\nu < n}$ of system~\eqref{eq:system-Lure-mr} described by the equations
\beq \label{eq:system-ROM}
\dot{\hat{x}} = \hat{A}\hat{x} +\hat{B}_u  \hat{u}   +\hat{B}_w \hat{w}, \, \hat{y} = \hat{C}_y \hat{x},  \, \hat{z} = \hat{C}_z \hat{x},  \, \hat{w} = - \varphi(\hat{z}),  \! \!
\eeq
with ${\hat{x}\in\R^\nu}$, ${ \hat{u} \in\R}$, ${\hat{y}\in\R}$,  ${\hat{w}\in\R}$, ${\hat{z}\in\R}$, ${\hat{A}\in\R^{\nu \times \nu}}$, ${\hat{B}_u\in\R^{\nu \times 1}}$, ${\hat{B}_w\in\R^{\nu \times 1}}$, ${\hat{C}_y\in\R^{1\times \nu}}$, and ${\hat{C}_z\in\R^{1\times \nu}}$. Let 
\beq \nn
\hat{G}(s) 
=
\hat{C} (sI-\hat{A})^{-1} \hat{B}
= 
\bma
\begin{array}{cc}
\hat{G}_{yu}(s) & \hat{G}_{yw}(s)\\
\hat{G}_{zu}(s) & \hat{G}_{zw}(s)
\end{array}
\ema
,
\eeq
in which ${\hat{B} = [\, \hat{B}_{u} \ \hat{B}_{w} \,]}$ and  ${\hat{C} = [\, \hat{C}_{y}^{\transpose} \ \hat{C}_{z}^{\transpose} \,]^{\transpose}}$, respectively.  
The transfer function of the error system is defined as
\beq \nn
\tilde{G}(s) = G(s) - \hat{G}(s). 
\eeq
The reduced order model~\eqref{eq:system-ROM} is constructed by replacing the linear dynamics of the original system
\beq \label{eq:system-Lure-mr-linear}
\dot{x} = Ax +B_u u +B_w w, \ y = C_y x,  \ z = C_z x, 
\eeq
with a linear system described by the equations
\beq \label{eq:system-ROM-linear}
\dot{\hat{x}} = \hat{A}\hat{x} +\hat{B}_u  \hat{u}   +\hat{B}_w \hat{w}, \ \hat{y} = \hat{C}_y \hat{x},  \ \hat{z} = \hat{C}_z \hat{x}, 
\eeq
and by leaving the static nonlinearity ${\varphi}$ unchanged, as illustrated in Fig.~\ref{fig:model_reduction}.

This work addresses basic, yet open, questions about the \textit{global} behavior of the reduced order model~\eqref{eq:system-ROM}. For example, 
suppose that the original system~\eqref{eq:system-Lure-mr} is multistable or oscillatory. Further, suppose an error bound is available for the approximation of the linear dynamics. One can expect that~\eqref{eq:system-ROM} is indeed a good reduced order model if the error is small, but how does one \textit{quantify} this? Can one construct  better  reduced order models in a systematic and computationally appealing fashion? Is it sensible to use the reduced order model for analysis and design?

The paper~\citep{besselink2009error} has shown that these questions can be addressed for (incrementally) \textit{stable} Lure systems using sector conditions and small gain conditions.
The next sections show that similar conclusions can be drawn for \textit{dominant} Lure systems by combining existing linear model reductions methods with dominance theory.

\section{Dominance theory} \label{sec:preliminaries}

Consider a continuous-time, nonlinear, time-invariant system and its linearization described by the equations
\begin{align}
\,\Sigma : \quad \ \dot{x} &= f(x) +Bu, \quad \quad \quad \ \ \ y = \ Cx , \label{eq:system-nonlinear} \\
\delta\Sigma :\quad \dot{\delta x} &= \partial f(x) \delta x + B \delta u, \quad ~\, \delta y =~ C \delta x,   \label{eq:system-nonlinear-linearization}
\end{align}
in which ${x\in\R^n}$, ${u\in\R^m}$,  ${y\in\R^l}$,  
${f:\R^{n} \to \R^{n}}$ is a continuously differentiable vector field,   ${B\in\R^{n\times m}}$,   and ${C\in\R^{m\times n}}$ are constant matrices, ${\delta x\in\R^n}$, ${\delta u\in\R^m}$,   ${\delta y\in\R^l}$   (identified with the respective tangent spaces), and ${\partial f}$ is the Jacobian of the vector field $f$.

\begin{definition} \label{def:dominance}
The system~\eqref{eq:system-nonlinear} is $p$-dominant with rate ${\lambda \in \Rge}$ (for ${u=0}$) if there exist ${\varepsilon \in \Rge}$ and a symmetric matrix ${P\in\R^{n\times n}}$, with inertia\footnote{The inertia of the matrix ${A \in \R^{n \times n}}$ is defined as $\ine(A)=(n_{-},n_{0},n_{+})$, where $n_{-}$ is the number of eigenvalues of $A$ in the open left half-plane, $n_{0}$ is the number of eigenvalues of $A$ on the imaginary axis, and $n_{+}$ is the number of eigenvalues of $A$ in the open right half-plane, respectively.  
} $\ine(P) = ( p, 0,n-p)$, such that the prolonged system~\eqref{eq:system-nonlinear}-\eqref{eq:system-nonlinear-linearization}
satisfies
\beq  \label{eq:dominance-LMI}
\bma
\begin{array}{c}
 \dot{\delta x}  \\
\delta x
\end{array}
\ema^{\transpose}
\bma
\begin{array}{cc}
0 & P \\
P &  2\lambda P + \varepsilon I
\end{array}
\ema
\bma
\begin{array}{c}
\dot{\delta x} \\
\delta x
\end{array}
\ema \le 0 
\eeq 
for every $(x,\delta x) \in \R^n \times \R^n$. The property is strict if ${\varepsilon >}0$.
\end{definition}

The property of $p$-dominance ensures the existence of a splitting between $p$ dominant modes and $n-p$ transient modes.  In particular, the matrix ${\partial f}(x) +\lambda I$ has $p$ unstable eigenvalues and $n-p$ stable eigenvalues for every ${x\in\R^n}$. For linear, time-invariant systems, such eigenvalues are referred to as \textit{dominant and non-dominant}, respectively.

\begin{definition} \label{def:pgain}
The system~\eqref{eq:system-nonlinear} is said to have \textit{finite (differential) ${\L_{2,p}}$-gain (from $u$ to $y$) less than ${\gamma \in \Rge}$ with rate ${\lambda \in \Rge}$} if there exist ${\varepsilon \in \Rge}$ and a symmetric matrix ${P\in\R^{n\times n}}$, with inertia   $(p, 0, n-p)$,   such that the conic constraint
\beq \label{eq:system-linear-MIMO-conic-IO}
\begingroup 
\setlength\arraycolsep{1.3pt}
\bma
\begin{array}{c}
\dot{\delta x} \\
\delta x
\end{array}
\ema^{\transpose} \!
\bma
\begin{array}{cc}
0 & P \\
P &  2\lambda P + \varepsilon I
\end{array}
\ema \!
\bma
\begin{array}{c}
  \dot{\delta x} \\
\delta x
\end{array}
\ema 
\le
\bma
\begin{array}{c}
\delta y \\
\delta u
\end{array}
\ema^{\transpose} \!
\bma
	\begin{array}{cc}
	-I & 0 \\
	 0 & \gamma^2 I
	\end{array}
	\ema \!
\bma
\begin{array}{c}
\delta y \\
\delta u
\end{array}
\ema  \! \! 
\endgroup
\eeq
holds along the solutions of the prolonged system~\eqref{eq:system-nonlinear}-\eqref{eq:system-nonlinear-linearization}.
The \textit{(differential) $\L_{2,p}$-gain of system~\eqref{eq:system-nonlinear}  (from $u$ to $y$) with rate $\lambda$} is defined as ${\gamma_{\lambda} = \inf \left\{ \gamma \in \Rge  :  \text{\eqref{eq:system-linear-MIMO-conic-IO} holds} \right\}}.$ The properties are strict if $\varepsilon >0$. 
\end{definition}

The asymptotic behavior of a $p$-dominant system is $p$-dimensional and, hence, its attractors
are severely constrained for small values of $p$~\citep{forni2018differential}.

\begin{theorem}  \label{thm:asymptotic}
Assume system~\eqref{eq:system-nonlinear} is strictly $p$-dominant with rate ${\lambda \in \Rge}$. Then every bounded solution of~\eqref{eq:system-nonlinear} converges asymptotically to
\begin{description}
\item[{\small $\bullet$}] the unique equilibrium  point if ${p=0}$; 
\item[{\small $\bullet$}] a (possibly non-unique) equilibrium  point if ${p=1}$;
\item[{\small $\bullet$}] a simple attractor if ${p=2}$, \textit{i.e.} an equilibrium  point, a set of equilibrium  points and their connected arcs or a limit cycle.
\end{description} 
\end{theorem} 

The dominance properties of an interconnected system can be studied
using small ${\L_{2,p}}$-gain conditions~\citep{forni2018differential,padoan2019norm}.

\begin{theorem}[Small-gain theorem for $p$-dominance] \label{thm:small-gain}
Let $\Sigma_i$ be a system with input ${u_i\in\R^{m_i}}$, output ${y_i\in\R^{l_i}}$, and (strict) $\L_{2,p_i}$-gain less than ${\gamma_i\in\Rge}$ with rate ${\lambda \in \Rge}$, with ${i \in \{1,2\}}$. Then the closed-loop system defined by the feedback interconnection equations  ${u_1 = -y_2}$ and ${u_2 = y_1}$  is strictly ${(p_1+p_2)}$-dominant with rate $\lambda$ if ${\gamma_1 \gamma_2 <1}$. 
\end{theorem}

\subsection{Dominant systems in the frequency domain}

Consider a continuous-time, linear, time-invariant system described by the equations 
\beq \label{eq:system-linear-MIMO}
\quad \dot{x} = Ax+Bu, \quad y=Cx,
\eeq
in which ${x \in\R^n}$, ${u\in\R^m}$, ${y\in\R^l}$, and ${A\in\R^{n \times n}}$, ${B\in\R^{n \times m}}$, and ${C\in\R^{l\times n}}$, are constant matrices, respectively. Let $G(s) = C(sI-A)^{-1} B$ be its transfer function and, given ${\lambda \in \Rge}$, let ${G_\lambda(s) = G (s-\lambda)}$ be its shifted transfer function, provided $G$ is well-defined on the axis ${\Re(s) = - \lambda}$.

The ${\L_{2,p}}$ gain is intimately connected to the \textit{${\H_{\infty,p}}$ norm} of the transfer function of system~\eqref{eq:system-linear-MIMO}, defined as
\beq \label{eq:Hq-norm-lambda-strip} 
\norm{G}_{\infty,p} = \displaystyle \sup_{\omega \in  \R } {\sigma_{\max}} (G_{\lambda}(i\omega)) , 
\eeq 
where ${\sigma_{\max}}(M)$ is the largest singular value of the matrix ${M\in\Co^{l\times m}}$ and ${G}$ has $p$ poles in the open half-plane ${\Re(s)>-\lambda}$, respectively. The ${\H_{\infty,p}}$ norm is not uniquely defined by the integer $p$, as it depends on the parameter $\lambda$.

\begin{theorem} \citep{padoan2019norm} \label{thm:pgain}
Assume system~\eqref{eq:system-linear-MIMO} has (strict) $\L_{2,p}$-gain  ${\gamma_{\lambda} }$ with rate ${\lambda \in \Rge}$. Then ${\gamma_{\lambda}  =  \norm{G}_{\infty,p}.}$
\end{theorem}

The property of $p$-dominance can be also characterized \textit{graphically} in the frequency domain~\citep{felix2018analysis}.  Consider a single-input, single-output, Lure system described by the equations 
\beq \label{eq:system-Lure}
\dot{x} = Ax +Bu, \quad y = Cx,  \quad u =  -   \varphi(y),
\eeq
in which ${x\in\R^n}$, ${u \in\R}$, ${y \in\R}$, ${A\in\R^{n \times n}}$, ${B\in\R^{n \times 1}}$, and ${C\in\R^{1\times n}}$  are constant matrices, and ${\varphi:\R \to \R}$ is a continuously differentiable function which satisfies 
the \textit{differential sector condition} ${\partial \varphi \in  [\alpha, \beta]}$, defined as
\beq \label{eq:sector-condition} 
(\partial \varphi (y) \delta y - \alpha \delta y) (\partial \varphi (y) \delta y - \beta \delta y) \leq 0, \  \forall \, y \in \R,
\eeq
with ${-\infty \le \alpha < \beta \le \infty}$. Let 
${G(s) = C(sI-A)^{-1}B}$.

\begin{theorem}[Circle criterion for $p$-dominance] \label{thm:circle_criterion}
Consider \linebreak system~\eqref{eq:system-Lure} and let $\lambda \in \Rge$. Assume 
$\partial \varphi \in [\alpha, \beta]$, $G_{\lambda}$ has $q$ unstable poles and no poles along the imaginary axis,
the Nyquist diagram of $G_{\lambda}$ encircles ${(p-q)}$ times the point $-\tfrac{1}{\alpha}$  in the clockwise direction, and one of the following  mutually exclusive   conditions hold\footnote{$D(\alpha, \beta)$  denotes the closed disk in the complex plane associated with the sector $[\alpha, \beta]$, with ${\alpha < \beta}$. The notation is standard and is defined, \textit{e.g.}, in~\citep[p.82]{felix2018analysis}.}
\begin{itemize}
\item[(a)] ${\alpha \beta \ge 0}$ and the Nyquist diagram of $G_{\lambda}$ lies \textit{outside} the disk $D(\alpha, \beta)$.
\item[(b)] ${\alpha \beta <0}$ and the Nyquist diagram of $G_{\lambda}$ lies \textit{inside} the disk $D(\alpha, \beta)$.
\end{itemize}
Then system~\eqref{eq:system-Lure} is strictly $p$-dominant with rate $\lambda$.
\end{theorem}

\section{Main results} \label{sec:main-results}

\begin{figure}[h!]
\centering
\begin{tikzpicture}[scale=.75, black,every node/.style={transform shape}]
      
\draw[line width = .5 pt]  (10.6875,-0.375) rectangle (11.6875,-1.375);
\node at (11.1875,-0.875) {$\hat{G}_{zw}^{-}$};
\draw[rounded corners,line width = .5 pt]  (10.6875,-2.6964) rectangle (11.6875,-3.6964);
\node at (11.1875,-3.1964) {$-\varphi(\cdot)$};
\draw[-latex, line width = .5 pt] (12.1875,-2.0625)   -- (12.1875,-3.1964) -- (11.6875,-3.1964);
\draw[-latex, line width = .5 pt] (10.6875,-3.1964) -- (10.1875,-3.1964) -- (10.1875,-2) -- (10.6875,-2) ;
\draw [-latex, line width = .5 pt] (10.1875,-0.875) --(10.1875,0.4466) -- (10.6875,0.4466);
\draw [-latex,line width = .5 pt](11.6875,0.4466) -- (12.1875,0.4466)  -- (12.1875,-0.8194);
\draw[fill = white] (12.1875,-0.875) circle [radius=0.05];
\draw[semithick, dashed] (10,1.0716) rectangle (12.6875,-0.1159);
\draw[semithick, dashed] (9.9992,-0.3) rectangle (12.6883,-2.6076);
\draw  (10.6875,0.9466) rectangle (11.6875,-0.0534);
\node at (11.1875,0.4466) {$\tilde{G}_{zw}^{-}$};
\draw[-latex, line width = .5 pt] (11.6875,-0.875) -- (12.1319,-0.875); 
\draw[-latex, line width = .5 pt]  (10.6875,-1.5) rectangle (11.6875,-2.5);
\node at (11.1875,-2) {${G_{zw}^+}$};
\draw [-latex, line width = .5 pt](10.1875,-2) -- (10.1875,-0.875) -- (10.6875,-0.875);
\draw[fill = white] (12.1875,-2) circle [radius=0.05];
  \draw [-latex, line width = .5 pt](12.1875,-0.9375) -- (12.1875,-1.9375);
\draw [-latex, line width = .5 pt](11.6875,-2) -- (12.125,-2);
\node at (12.4375,0.8216) {$\tilde{G}$};
\node at (12.4375,-2.2951) {$\hat{G}$};

\draw [thick, -latex](12.875,-1.25) -- (15.25,-1.25);
\node at (14.0625,-0.75) { Model reduction};
  
\draw[line width = .5 pt]  (16.1875,-0.375) rectangle (17.1875,-1.375);
\node at (16.6875,-0.875) {$\hat{G}_{zw}^{-}$};
\draw[rounded corners,line width = .5 pt]  (16.1875,-2.6964) rectangle (17.1875,-3.6964);
\node at (16.6875,-3.1964) {$-\varphi(\cdot)$};
\draw[-latex, line width = .5 pt] (17.6875,-2.0625)   -- (17.6875,-3.1964) -- (17.1875,-3.1964);
\draw[-latex, line width = .5 pt] (16.1875,-3.1964) -- (15.6875,-3.1964) -- (15.6875,-2) -- (16.1875,-2) ;

\draw[semithick, dashed] (15.4992,-0.3) rectangle (18.1883,-2.6076);
  
\draw[-latex, line width = .5 pt]  (16.1875,-1.5) rectangle (17.1875,-2.5);
\node at (16.6875,-2) {${G_{zw}^+}$};
\draw [-latex, line width = .5 pt](15.6875,-2) -- (15.6875,-0.875) -- (16.1875,-0.875);
\draw[fill = white] (17.6875,-2) circle [radius=0.05];
  \draw [-latex, line width = .5 pt](17.1875,-0.875) -- (17.6875,-0.8661) -- (17.6875,-1.9375);
\draw [-latex, line width = .5 pt](17.1875,-2) -- (17.625,-2);
 
\node at (17.9375,-2.2951) {$\hat{G}$};

\end{tikzpicture} 
\centering
\caption{Diagrammatic illustration of the model reduction method described in Algorithm~\ref{alg:mr-Lure}.
}
\label{fig:model_reduction_detail_new}
\end{figure}
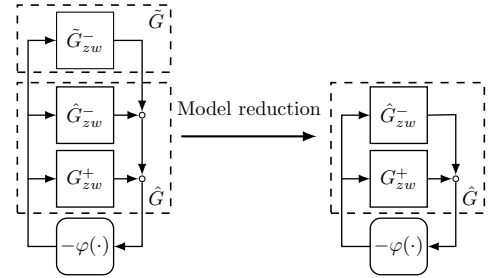

The first step of our model reduction approach for dominant Lure systems is to split the linear dynamics into dominant and non-dominant components. Standard model reduction methods are then applied to the non-dominant component, as illustrated in Fig.~\ref{fig:model_reduction_detail_new}. The main steps are summarized in Algorithm~\ref{alg:mr-Lure}.  This approach preserves by construction the dominance properties of the linear dynamics of the original system. Furthermore, taking advantage of small-gain conditions for $p$-dominance, it guarantees that the behavior of the original original system is well captured by that of the reduced order model, when the approximation error is small. This is formalized in Theorem~\ref{thm:1}.
 
\begin{algorithm}[h!] 
\algnewcommand\algorithmicto{\textbf{to}}
\algnewcommand\algorithmicbreak{\textbf{break}}
\algnewcommand\algorithmicstop{\textbf{stop}}
\algnewcommand\algorithmicass{\textbf{Assumption}}
\renewcommand{\algorithmicrequire}{\textbf{Input}}
\renewcommand{\algorithmicensure}{\textbf{Output}}
\caption{}
 \label{alg:mr-Lure}  
\algorithmicrequire{: The system~\eqref{eq:system-Lure-mr}}\\
\algorithmicensure{: The reduced order model~\eqref{eq:system-ROM}}\\
\algorithmicass{: {Strict\,$p$-dominance\,with\,rate\,$\lambda$\,of\,system\,\eqref{eq:system-Lure-mr}}} 
\begin{algorithmic}[1] 
\State Consider the linear dynamics~\eqref{eq:system-Lure-mr-linear} and, upon a possible change of coordinates, let
\beq  \nn
A =
\bma
\begin{array}{cc}
A^+ & 0 \\
0 & A^-
\end{array}
\ema, \
B =
\bma
\begin{array}{c}
B^+ \\
B^-
\end{array}
\ema, \
C =
\bma
\begin{array}{cc}
C^+  &
C^- 
\end{array}
\ema, 
\eeq
where ${\spectrum{A^+}}$ and ${\spectrum{A^-}}$ contain $p$ dominant eigenvalues and $n-p$ non-dominant eigenvalues,  respectively.  
\State Construct a reduced order model 
\beq \nn
\dot{\hat{x}}^- = \hat{A}^-\hat{x}^- +\hat{B}^-  \hat{v}^-, \quad \hat{y}^- = \hat{C}^- \hat{x}^-, 
\eeq
of the non-dominant linear dynamics 
\beq \nn
\dot{x}^- = {A}^- {x}^- + {B}^-{v}^-, \quad {y}^- = {C}^- {x}^-, 
\eeq
using a stability-preserving model reduction method. 
\State Define the reduced order model~\eqref{eq:system-ROM} as
\beq  \nn
\hat{A} =
\bma
\begin{array}{cc}
A^+ & 0 \\
0 & \hat{A}^-
\end{array}
\ema, \
B =
\bma
\begin{array}{c}
B^+ \\
\hat{B}^-
\end{array}
\ema, \
C =
\bma
\begin{array}{cc}
C^+  &
\hat{C}^- 
\end{array}
\ema .
\eeq 
\end{algorithmic}
\end{algorithm}

\begin{theorem} \label{thm:1}
Consider system~\eqref{eq:system-Lure-mr} and a reduced order model~\eqref{eq:system-ROM} constructed as  in Algorithm~\ref{alg:mr-Lure}.  Assume
\begin{description}
\item[(A1)]  \eqref{eq:system-Lure-mr-linear} is strictly   $p$-dominant with rate ${\lambda\in\Rge}$,
\item[(A2)] ${\partial \varphi \in  [-\mu, \mu]}$, with  ${\mu\in \Rge}$,
\item[(A3)] ${\norm{\tilde{G}}_{\infty,0} \le \epsilon}$, with ${0<\epsilon<\mu^{-1}}$,
\item[(A4)] ${\norm{\hat{G}_{zw}}_{\infty,p} < \mu^{-1} - \epsilon}$.
\end{description}
Then system~\eqref{eq:system-Lure-mr} is strictly $p$-dominant with rate ${\lambda}$.
\end{theorem}

\begin{pf}
By the circle criterion for $p$-dominance, system~\eqref{eq:system-Lure-mr} is $p$-dominant with rate ${\lambda}$ if 
\beq \label{eq:thm1-proof-1}
\norm{G_{zw}}_{\infty,p} < \mu^{-1},
\eeq
since, by assumptions (A1) and (A2), \eqref{eq:system-Lure-mr} is $p$-dominant with rate ${\lambda}$ (and, thus, the transfer function ${G(s-\lambda)}$ has $p$ unstable poles and no poles on the imaginary axis) and ${\partial \varphi \in  [-\mu, \mu]}$. However, \eqref{eq:thm1-proof-1} follows directly from assumptions (A3) and (A4), as
\begin{align*}
\norm{G_{zw}}_{\infty,p}
\stackrel{\text{def}}{=} \norm{\hat{G}_{zw}  + \tilde{G}_{zw}}_{\infty,p} 
\stackrel{(\text{A3})}{\le} \norm{\hat{G}_{zw}}_{\infty,p} + \epsilon  
\stackrel{(\text{A4})}{<} \mu^{-1}.
\end{align*} %~\hspace*{\fill}~$\blacksquare$
\end{pf}

Theorem~\ref{thm:1} establishes that the dominance properties of the 
original closed-loop system can be inferred from those of the reduced order model if the approximation error between the original linear component $G$ and the reduced linear component $\hat{G}$ is sufficiently small.  
From Algorithm~\ref{alg:mr-Lure}, by construction, Assumption (A1) guarantees that the reduced linear dynamics $\hat{G}$ is $p$-dominant with rate $\lambda$. 
Assumption (A2) and the circle criterion for $p$-dominance 
ensure that the closed-loop reduced order model is also $p$-dominant 
with the same rate  if ${\norm{\hat{G}_{zw}}_{\infty,p} < \mu^{-1} }$ (which, in turn, is implied by assumption (A4)). The approximation error bound of assumption (A3) expresses the fact that the  linear   dynamics of the reduced order model is ``close'' to that of the original system in the ${\H_{\infty,0}}$ norm.
Finally, the small ${\LTp}$ gain condition of assumption (A4) ensures that the dominance properties of the original system are captured by those of the the reduced order model when the approximation error is regarded as a perturbation.

\begin{remark}
Theorem~\ref{thm:1} may be conservative as it requires only minimal information about the static nonlinearity $\varphi$; it applies to \textit{any} static nonlinearity $\varphi$ such that ${\varphi(0) =0}$ and ${\partial \varphi \in  [-\mu, \mu]}$. The model reduction framework is therefore robust to nonlinear uncertainty.
\hspace*{\fill} $\blacktriangle$
\end{remark}

\begin{figure}[h!]
\centering
\usetikzlibrary{arrows.meta}
\tikzset{>={Latex[width=3pt,length=3pt]}}
\usetikzlibrary{decorations.pathreplacing}
\begin{tikzpicture}[black, scale=.7, every node/.style={transform shape}]

\draw[line width = .5 pt]  (11.8125,3.3125) rectangle (12.8125,2.3125);
\node at (12.3125,2.8125) {$\hat{G}_{zw}^{-}$};
\draw[rounded corners,line width = .5 pt]  (11.8125,0.9375) rectangle (12.8125,-0.0625);
\node at (12.3125,0.4375) {$-\varphi(\cdot)$};
\draw[-latex, line width = .5 pt] (13.3125,1.625)   -- (13.3125,0.4375) -- (12.8125,0.4375);
\draw[-latex, line width = .5 pt] (11.8125,0.4375) -- (11.3125,0.4375) -- (11.3125,1.6875) -- (11.8125,1.6875) ;
\draw [-latex, line width = .5 pt] (11.3125,2.8125) --(11.3125,4.5625) -- (11.8125,4.5625);
\node at (11.125,3.6875) {$w$};
\node at (13.5625,3.6875) {$\tilde{z}$};
\draw [-latex,line width = .5 pt](12.8125,4.5625) -- (13.3125,4.5625)  -- (13.3125,2.8681);
\draw[fill = white] (13.3125,2.8125) circle [radius=0.05];
\draw[semithick, dashed] (11.125,5.1875) rectangle (13.8125,4);
\node at (15,4.5625) {{ Gain $\tilde{\gamma}$}};
\draw[semithick, dashed] (11.1242,3.3875) rectangle (13.8133,1.0625);
\node at (15,1.75) {{ Gain ${\hat{\gamma}}$}};
\draw  (11.8125,5.0625) rectangle (12.8125,4.0625);
\node at (12.3125,4.5625) {$\tilde{G}_{zw}^{-}$};
\draw[-latex, line width = .5 pt] (12.8125,2.8125) -- (13.2569,2.8125); 
\draw[-latex, line width = .5 pt]  (11.8125,2.1875) rectangle (12.8125,1.1875);
\node at (12.3125,1.6875) {${G_{zw}^+}$};
\draw [-latex, line width = .5 pt](11.3125,1.6875) -- (11.3125,2.8125) -- (11.8125,2.8125);
\draw[fill = white] (13.3125,1.6875) circle [radius=0.05];
  \draw [-latex, line width = .5 pt](13.3125,2.75) -- (13.3125,1.75);
\draw [-latex, line width = .5 pt](12.8125,1.6875) -- (13.25,1.6875);
\node at (13.5625,4.9375) {$\tilde{G}$};
\node at (13.5625,1.375) {$\hat{G}$};

\draw [-latex](18.0625,0.125) -- (18.0625,5);
\draw [-latex](16.0625,2.5) -- (21,2.5);
\draw[thick] plot[smooth cycle, tension=.7] coordinates {(18.6733,2.5) (18.5069,2.8888) (18.0625,2.9444) (17.8405,2.5832) (18.0625,2.4168) (18.2289,2.4168)(18.2289,2.5832) (18.0625,2.5832) (17.8405,2.4168) (18.0625,2.0556) (18.5069,2.1112)} ;
\node at (20.5625,2.125) {{ $\mathbf{Re} \,\hat{G}_{\lambda}(i\omega)$}};
\node at (17.0625,4.75) {{ $\mathbf{Im} \,\hat{G}_{\lambda}(i\omega)$}};
\node at (16.1875,2.125) {{\large $-\tfrac{1}{\mu}$}}; 
\node at (19.25,3.75) {{\large${\mathbf{{\epsilon}}} $}}; 

\begin{scope}[even odd rule, shift={(2.0625,0)}] 
\clip (16,2.5) ellipse (1.5 and 1.5) ellipse (1.0 and 1.0);
\draw[thick, pattern=north west lines,  pattern color=gray!50]  (16,2.5) ellipse (1.9 and 1.9);
\end{scope} 
\draw[semithick,color=gray!80 ]  (18.0625,2.5) ellipse (1.0 and 1.0);
\draw[semithick,color=gray!80 ]  (18.0625,2.5) ellipse (1.5 and 1.5);
\draw[<->,thick] (18.7301,3.2232) -- (19.0901,3.5832);

\draw [thick, decorate, decoration={brace, amplitude=5pt}](14.0625,5.1875) -- (14.0625,4);
\draw [thick, decorate, decoration={brace, amplitude=5pt}](14.0625,3.5) -- (14.0625,0);
\end{tikzpicture} 
\centering
\caption{Theorem~\ref{thm:1}: a small ${\LTp}$ gain intepretation (left) and a graphical interpretation (right), respectively.
}
\label{fig:theorem_new}
\end{figure}
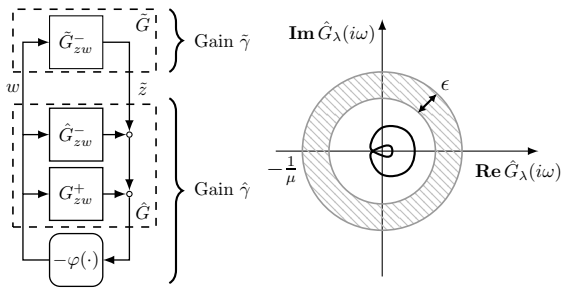

Theorem~\ref{thm:1} can be seen an application the small ${\LTp}$ gain theorem.  Assumptions (A3) and (A4) imply the small ${\LTp}$ condition ${\gamma_1 \gamma_2 < 1}$, in which ${\tilde{\gamma} = \norm{\tilde{G} }_{\infty,0}}$ and ${\hat{\gamma} = (\mu^{-1} - \norm{\hat{G}}_{\infty,p} )^{-1}}.$ This guarantees that the closed-loop system in Fig.~\ref{fig:theorem_new} (left) is $p$-dominant, as the product of the differential  gain  ${\tilde{\gamma}}$ (from $w$ to $\tilde{z}$) and the differential gain ${\hat{\gamma}}$ (from $\tilde{z}$ and $w$) is strictly less than one. \hspace*{\fill}

The small ${\LTp}$ condition (A4)    admits a nice graphical interpretation. From a geometrical viewpoint, it implies that the distance between the critical circle and the Nyquist diagram of the shifted transfer function ${\hat{G}_{\lambda}}$ must be at least ${\epsilon}$, as illustrated in Fig.~\ref{fig:theorem_new} (right). This implies that $D(-\hat{\mu}, \hat{\mu})$, with ${\hat{\mu} = (\mu^{-1} - \epsilon)^{-1}}$, is a $p$-disk margin for the reduced order model~\citep{padoan2019feedback}. This suggests that claim (i) can be established by a homotopy argument  (given, \textit{e.g.}, in   \cite[Theorem 3]{glover1983robust}).

A reduced order model can be very useful for analysis. As such, it should be able to reproduce faithfully the behavior of the original system. This point  is  addressed by the next result, which provides conditions under which the dominance properties of the original system are preserved by the reduced order model. The proof is omitted as it closely parallels that of Theorem~\ref{thm:1}, with  ${G}$ and ${-\tilde{G}}$ playing the roles of ${\hat{G}}$ and ${\tilde{G}}$, respectively. 

\begin{corollary} \label{corollary:1}
Consider system~\eqref{eq:system-Lure-mr} and  a  reduced order model~\eqref{eq:system-ROM}
constructed as in Algorithm~\ref{alg:mr-Lure}. Assume
\begin{description}
\item[(A1)$^{\star}$]\eqref{eq:system-ROM-linear} is  strictly   $p$-dominant with rate ${\lambda\in\Rge}$,
\item[(A2)$^{\star}$] ${\partial \varphi \in  [-\mu, \mu]}$, with  ${\mu\in \Rge}$,
\item[(A3)$^{\star}$] ${\norm{\tilde{G}}_{\infty,0} \le \epsilon}$, with ${0<\epsilon<\mu^{-1}}$;
\item[(A4)$^{\star}$] ${\norm{G_{zw}}_{\infty,p} < \mu^{-1} - \epsilon}$.
\end{description}
Then the reduced order model~\eqref{eq:system-ROM} is strictly $p$-dominant with rate ${\lambda}$.
\end{corollary}

\section{Balanced truncation of Lure systems} \label{sec:mr}
 
The results developed in the previous section require a stable  reduced order model of the non-dominant linear dynamics   together with an \textit{a priori} error bound. In general, any model reduction method that produces a reduced order model with these properties can be used to reduce the original Lure system. 

Model reduction by balanced truncation~\citep[Section 7]{antoulas2005approximation} is an obvious candidate, as it preserves stability and provides an error bound that is easy to compute. We briefly recall this method for completeness.
Given a linear, time-invariant, minimal, stable system~\eqref{eq:system-linear-MIMO},  balancing consists in finding a coordinates transformation ${\bar{x} = T^{-1} x }$  such that the reachability gramian $P\in\R^{n \times n}$ and the observability gramian $Q\in\R^{n \times n}$ of the system, defined implicitly by the Lyapunov equations
\begin{align}
A P +PA^{\transpose} +BB^{\transpose} = 0, \\
A^{\transpose} Q +QA  +C^{\transpose}C = 0, 
\end{align} 
are both diagonal and equal. A \textit{balancing transformation} $T$ acts on the reachability and observability gramians as 
\beq
T^{-1} P T^{-\transpose}  = T^{\transpose} Q T =   \diag(\sigma_1, \ldots, \sigma_n), 
\eeq
in which case the corresponding realization is said to be \textit{(principal-axis) balanced}. The elements ${\sigma_1, \ldots, \sigma_n}$ are the \textit{Hankel singular values} of the system. These are system invariants  which  measure the  influence  of each state  on  the overall input-output behavior of the system.
Model reduction by balanced truncation consists in ordering the Hankel singular values and in eliminating ${n-\nu}$ state variables by truncation. The resulting reduced order model  is of order $\nu$, stable,   and satisfies the error bound
\beq
\norm{G - \hat{G}}_{\infty} \le 2 
\sum_{j=\nu+1}^{n} \sigma_{j} =\epsilon ,
\eeq
in which $G$ and $\hat{G}$ are the transfer functions of the original system and  of  the reduced order model, respectively.

As anticipated above, balanced truncation can be applied to  a  dominant Lure system as follows. First, one decomposes the linear dynamics of the original system~\eqref{eq:system-Lure} into the parallel interconnection of two subsystems described by the transfer functions $G^+$ and  $G^-$, which are strictly $p$-dominant and $0$-dominant with rate $\lambda$, respectively. Then a reduced order model of order $\nu^{-}$  of system ${G}^-$ is obtained by balanced truncation using the Lyapunov equations
\begin{align}
(A^{-}+\lambda I) P +P(A^{-}+\lambda I)^{\transpose} +B^{-}(B^{-})^{\transpose} = 0, \label{eq:Lyapunov-reachability}\\
(A^{-}+\lambda I)^{\transpose} Q +Q(A^{-}+\lambda I) +(C^{-})^{\transpose}C^{-} = 0, \label{eq:Lyapunov-observability}
\end{align} 
with $(A^{-}, B^{-}, C^{-})$ a minimal realization of order $n^{-}$ of ${G}^-$. The resulting reduced order model $\hat{G}^-$  is strictly $0$-dominant with rate $\lambda$ and satisfies the error bound
\beq \label{eq:error-dominant}
\norm{G^{-} - \hat{G}^{-}}_{\infty,0} \le 2 
\sum_{j=\nu^{-}+1}^{n^{-}} \sigma_{j} =\epsilon .
\eeq
Finally, a reduced order model is constructed by considering first the parallel interconnection of two subsystems $G^+$ and  $\hat{G}^-$ and then the feedback interconnection of the resulting system with the static nonlinearity $\varphi$, as illustrated in Fig.~\ref{fig:model_reduction_detail_new}. Balanced truncation allows for direct control over the error bound $\epsilon$ according to~\eqref{eq:error-dominant} by appropriately selecting the order $\nu^{-}$
of the reduced model $\hat{G}^-$.  In the context of dominant systems, the error bound $\epsilon$   is related to the rate $\lambda$ selected to compute the  $H_{\infty,0}$ norm. In principle, a careful selection of this additional parameter could lead to tighter reduction errors. This requires additional investigation.

\section{An illustrative example}  \label{sec:example}

The approximation of large-scale systems arising from the spatial discretization of the heat equation is a paradigmatic model reduction problem~\citep{chahlaoui2002collection,antoulas2005approximation}. The goal is often to capture the behavior of these systems around equilibrium. This example, instead, focuses on oscillatory regimes, motivated by the fact that the heat equation describes key physical phenomena involving transport, which play a fundamental role in fluid dynamics and neuroscience~\citep{keener1998mathematical}.

\begin{figure}[h!]
\centering
\usetikzlibrary{decorations.markings}
\usetikzlibrary{patterns}
\usetikzlibrary{arrows, arrows.meta}
\tikzset{>={Latex[width=3pt,length=3pt]}}
\begin{tikzpicture}[black, scale=.8, every node/.style={transform shape}]
\draw  (1.5,3) rectangle (2.5,2);
\node at (2,2.5) {$-k_P$};
\draw[rounded corners]  (3.5,3) rectangle (4.5,2);

\draw[ line width = 1 pt] (3.7,2.35) -- (3.85,2.35) -- (4.15,2.65) -- (4.35,2.65);
\draw[-latex, line width = 0.5 pt] (5.5556,2.5) -- (6.35,2.5);
\draw [-latex, line width = 0.5 pt](2.5,2.5) -- (3.5,2.5);
\draw [-latex, line width = 0.5 pt](9.5,2.5) -- (10.5,2.5) -- (10.5,1.5) -- (0.5,1.5) -- (0.5,2.5) -- (1.5,2.5);
\draw[fill = white] (5.5,2.5) circle [radius=0.05];
\draw [-latex, line width = 0.5 pt](5.5,3.5) -- (5.5,2.5556);
\node at (10.25,2.75) {$y =z$};
\node at (5.7776,3.1112) {$u$};
   
\draw[top color=white,bottom color=black!70] (6.5,28mm) arc (90:270:0.15 and 0.3)--++(3cm,0) arc (-90:-270:0.15 and 0.3)-- cycle;
\draw (9.5cm,28mm) arc (90:-90:0.15 and 0.3);
\draw (7cm,28mm) arc (90:270:0.15 and 0.3);
\draw (7.5cm,28mm) arc (90:270:0.15 and 0.3);
\draw (9cm,28mm) arc (90:270:0.15 and 0.3);
\node at (8.2,2.5) {$\cdots$};
 \node at (7.05,3) { $x_1$};
\node at (7.55,3) { $x_2$};
\node at (9.2,3) { $x_n$};
\node at (6.75,1.85) {\footnotesize $h$};
\draw [<->, thick](6.5,2) -- (7,2);
\draw [-latex, line width = 0.5 pt](4.5,2.5) -- (5.4444,2.5);
\node at (5,2.7776) {$w$};
\end{tikzpicture}
\centering
\caption{Heat flow control system for a homogeneous rod of unitary length via saturated proportional control.
}
\label{fig:setup}
\end{figure}
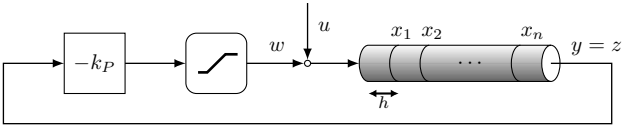

Consider the problem of regulating heat flow oscillations in a homogeneous rod of unitary length via saturated
proportional control, as illustrated in Fig.~\ref{fig:setup}.

The rod is insulated along its length and at the end. All temperature changes result from heat transfer at one end of the rod and by heat conduction along the rod.  By Fourier's law of heat conduction, the propagation of heat along the rod is described by the equation
\beq
\frac{\partial T}{\partial t} = \kappa \frac{\partial^2 T}{\partial \xi^2},
\eeq
in which ${T \in \Rge}$ is the temperature field on the rod, ${t \in \Rge}$ is the time variable, ${\xi \in (0,1)}$ is the spatial variable, and ${\kappa > 0}$ is the thermal diffusivity, respectively. The temperature is measured at the end of the rod, \textit{i.e.} ${y(t) = z(t) = T(1,t)}$. The temperature is initially zero and the heat flow source is placed at the beginning of the rod. The initial condition is
\beq
 T(\xi,0) = 0, \quad \xi \in (0,1)
\eeq
and the (Neumann) boundary conditions are
\beq
\frac{\partial T}{\partial \xi}(0,t) = u(t)+w(t) , \quad  \frac{\partial T}{\partial \xi}(1,t) = 0,
\eeq 
in which ${u \in \R}$ is the exogenous input and ${w \in \R}$ is defined as $w = -\sat(k_P z )$, where ${k_P>0}$ is the proportional gain and ${{\sat(z) = \tanh(z)}}$ is the saturation function.  The spatial domain is discretized into segments of length ${h = \tfrac{1}{n+1}}$ using the second order central difference scheme
\beq
\frac{\partial^2 T}{\partial z^2}(kh,t) \approx \frac{1}{h^2}(T_{k+1}(t) - 2T_{k}(t) + T_{k+1}(t)),
\eeq 
where ${T_{k}(t) = T(kh,t)}$. This yields a system described by the equations~\eqref{eq:system-Lure}, with 
${x = [\, T_{1} \ \cdots \ T_{n} \,]^{\transpose}}$, ${x(0) = 0}$, and 
\beq  \!
\scalebox{0.925}{$
\setlength{\arraycolsep}{1.25 pt}
\renewcommand{\arraystretch}{0.5}
A =
\tfrac{\kappa}{h^{2}}
\bma
\begin{array}{rrcrr} 
-1 &1 & & & \\
1 &-2 &\ddots & &\  \\
 &\ddots &\ddots & \ddots & \\ 
 & &\ddots &-2  &1 \\
 &  		 & &1  &-1 
\end{array}
\ema  \! , \, 
B=
\tfrac{\kappa}{h}
\bma
\begin{array}{cc} 
1 & 1\\
0 & 0 \\
\vdots & \vdots\\
0 & 0
\end{array}
\ema \! , \,
C=
\bma
\begin{array}{cc} 
0 & 0\\
0 & 0 \\
\vdots & \vdots\\
1 & 1
\end{array}
\ema^{\transpose} \!\!  . $} \!\!\! 
\eeq
The transfer function of the system is
\beq \label{eq:example-transfer-function}
G(s) 
= 
\bma
\begin{array}{cc}
G_{yu}(s) & G_{yw}(s)\\
G_{zu}(s) & G_{zw}(s)
\end{array}
\ema
=
\bgroup
\setlength{\arraycolsep}{3 pt}
\renewcommand{\arraystretch}{1}
\tfrac{\gamma_0}{\prod_{j=1}^n \left(s+\lambda_k\right)} 
\bma
\begin{array}{cc}
1 & 1 \\
1 & 1 
\end{array}
\ema
\egroup
,
\eeq
with ${\gamma_0 = \frac{ \kappa^n}{h^{2n-1}}}$ and ${\lambda_k = 2\tfrac{\kappa}{h^2} -2\tfrac{\kappa}{h^2}\cos\tfrac{(k-1)\pi}{n}}$ for every integer ${ k \in [1,n]}$. The static nonlinearity is ${\varphi(z) = \text{sat}(k_Pz)}$, which satisfies the differential sector condition ${\partial {\varphi} \in [0,k_P]}$.

The system captures the main features of the Goodwin model~\citep{murray2002mathematical},  a classical model in cellular physiology described by a cascade of reactions modelled by first-order lags and a negative feedback modelled by a Hill function. Thus, while approximating the heat equation away from equilibrium might not have significant engineering relevance, this could be crucial to understand, simulate and control biological rhythms.

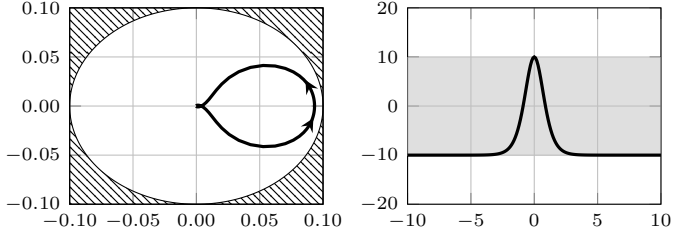
\begin{figure}[h!]
\centering 
\begin{tikzpicture}[black, every tick label/.append style={font=\scriptsize}]
\matrix{
\begin{axis}[
height = 0.23\textwidth,
width = 0.27\textwidth,
xmin=-.1,xmax=.1,
ymin=-.1,ymax=.1,
x tick label style={
        /pgf/number format/.cd,
            fixed,
            fixed zerofill,
            precision=2,
        /tikz/.cd
    },
y tick label style={
        /pgf/number format/.cd,
            fixed,
            fixed zerofill,
            precision=2,
        /tikz/.cd
    }    ,
grid
]
\draw[name path =A] (axis cs:-0.1,-0.1) -- (axis cs:-0.1,0.1) -- (axis cs:0.1,0.1) -- (axis cs:0.1,-0.1) -- cycle;
\addplot[name path =B,no markers, black] table [x index = {0}, y index = {1},  col sep=comma]{circle.csv};
\tikzfillbetween[of=A and B]{pattern=north west lines};  
\addplot [black, very thick,  %
postaction={decorate}, %
decoration={
  markings,
    mark=at position 0.5*\pgfdecoratedpathlength-7.5pt
  with {\arrow[xshift=2.5\pgflinewidth,>=stealth]{>}},
  mark=at position 0.5*\pgfdecoratedpathlength+7.5pt
  with {\arrow[xshift=2.5\pgflinewidth,>=stealth]{>}}
} %
]  table [x index = {0}, y index = {1},  col sep=comma]{nyquist_system.csv};
\end{axis}
~&~
\begin{axis}[
height = 0.23\textwidth,
width = 0.27\textwidth,
xmin=-10,xmax=10,
ymin=-20,ymax=20,
grid
]
\fill [color=gray!50,fill=gray!50,fill opacity=0.5] (axis cs:-10,-10) rectangle (axis cs:10,10);
\addplot[black, very thick]  table [x index = {0}, y index = {1},  col sep=comma]{sector_condition.csv};
\end{axis}
\\[1 pt]
};
\end{tikzpicture}
\vspace{-.5cm}
\centering
\caption{{Left}: Nyquist diagram of the shifted transfer function ${H_{zw}}$ defined by~\eqref{eq:example-transfer-function-loop} for ${\kappa = 1}$, ${n=29}$, ${k_P=20}$ and ${\lambda = 12}$. The disk $D(-\mu,\mu)$, with  ${\mu =  \tfrac{k_P}{2} =  10}$, is represented by diagonal lines. {Right}: Derivative of the static nonlinearity ${\bar{\varphi}}$. The differential sector condition ${ \partial \bar{\varphi} \in [-\mu,\mu]}$  is represented by the shaded area.
}
\label{fig:nyquist-P}
\end{figure}

To illustrate our model reduction framework and to ensure that assumptions (A1) and (A2) of Theorem~\ref{thm:1} hold, we consider the loop transformation defined as 
\beq \label{eq:loop-transformation}
\bar{\varphi}(z) = \varphi(z) - \tfrac{k_P}{2} z .
\eeq
This converts the differential sector condition ${\partial {\varphi} \in [0,k_P]}$ into ${\partial \bar{\varphi} \in [- \mu, \mu]}$, with  ${\mu = \tfrac{k_P}{2}}$, and the transfer function $G_{zw}$ into
\beq \label{eq:example-transfer-function-loop}
H_{zw}(s) = \frac{G_{zw}(s)}{1+ \tfrac{k_P}{2} G_{zw}(s)}.
\eeq 
The root locus of the transfer function $H_{zw}$ reveals that for any choice of the step size $h$ it is possible to establish strict $2$-dominance for ${k_P\in[\underline{k}_P,\overbar{k}_P]}$, with 
${0<\underline{k}_P <\overbar{k}_P}$. For example, let ${\kappa = 1}$, ${n=29}$, ${k_P=20}$  and ${\lambda = 12}$. Then $H_{zw}$ has exactly two unstable poles for ${k_P = 20}$.  Direct inspection of the Nyquist diagram and the circle criterion for $p$-dominance imply that the system is strictly $2$-dominant with the same rate, as illustrated in Fig.~\ref{fig:nyquist-P}. The system has therefore a unique limit cycle, since all solutions are bounded and the equilibrium at the origin is unstable.

\begin{figure}[t!]
\centering
\usetikzlibrary{spy}
\tikzset{new spy style/.style={spy scope={%
 magnification=5,
 size=1.25cm, 
 connect spies,
 every spy on node/.style={
   circle,
   draw,
   },
 every spy in node/.style={
   draw,
   circle,
   fill=gray!10,
   }
  }
 }
} 
\begin{tikzpicture}[new spy style,every tick label/.append style={font=\scriptsize}]
\begin{loglogaxis}[
height = 0.225\textwidth,
width = \columnwidth,
xmin=1e-0, xmax=1e03,
ymin=1e-6, ymax=1e0,
xlabel={{$\omega$} \textrm{[rad/s]}},          
xlabel near ticks,
grid
]
\node[left]  			at (axis cs: 9*1e2,3*1e-2) {{\footnotesize ${\tfrac{1}{\mu}=0.1}$}};
\node[left, rotate=-10]  at (axis cs: 9*1e2,6*1e-4) {{\footnotesize ${\nu=3}$}};
\node[left, rotate=-11]  at (axis cs: 9*1e2,7*1e-5) {{\footnotesize ${\nu=4}$}};
\node[left, rotate=-12]  at (axis cs: 9*1e2,8*1e-6) {{\footnotesize ${\nu=5}$}};

\addplot [black, thick]  table [x index = {0}, y index = {1},  col sep=comma]{bode_system.csv};
\addplot [black, thick,  dashed]  table [x index = {0}, y index = {1},  col sep=comma]{bode_rom1.csv};
\addplot [black, thick, dashdotted]  table [x index = {0}, y index = {1},  col sep=comma]{bode_rom2.csv};
\addplot [black, thick, densely dotted]  table [x index = {0}, y index = {1},  col sep=comma]{bode_rom3.csv};
\addplot [blue, thick ]  table [x index = {0}, y index = {1},  col sep=comma]{bode_mu_bound.csv};
\end{loglogaxis}
\spy[size=1.5cm,magnification=7] on (0.125,2.1) in node at (1.5,0.9);
\end{tikzpicture}
\vspace{-.25cm}
\centering
\caption{Magnitude  of the frequency response of the transfer function ${H_{zw}(s -\lambda)}$  defined by~\eqref{eq:example-transfer-function-loop},  with ${\kappa = 1}$, ${n=29}$, ${k_P=20}$, ${\lambda = 12}$, and   the frequency response of the transfer function ${\hat{H}_{zw}(s -\lambda)}$  of the reduced order model obtained via balanced truncation. 
}
\label{fig:bode_bt}
\end{figure}
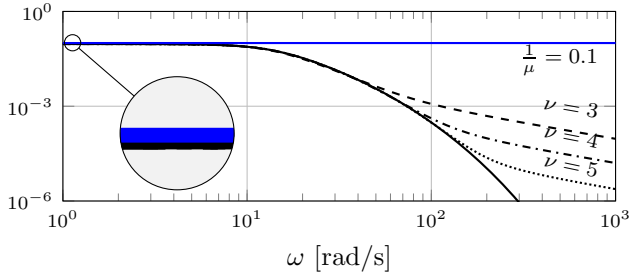

Simulations have been run to assess the properties of different reduced order models. Algorithm~\ref{alg:mr-Lure} and classical balanced truncation have been used to obtain reduced order models of the original system of order ${\nu = 3, 4, 5}$.  Assumption (A3) of Theorem~\ref{thm:1} holds, as the linear dynamics of the resulting reduced models are such that ${\norm{\tilde{H}_{zw}}_{\infty,0} \le \epsilon_{\nu}}$, with ${\epsilon_3 = 3.27\cdot 10^{-3}}$,    ${\epsilon_4 = 2.55\cdot10^{-4}}$, ${\epsilon_5 = 2.28\cdot10^{-5}}$, indicating that the approximation error ${\epsilon_{\nu}}$ is a strictly decreasing function of $\nu$.  Figure~\ref{fig:bode_bt} shows the magnitude of the shifted frequency response of the transfer function ${H_{zw}}$ defined by~\eqref{eq:example-transfer-function-loop}, with ${\kappa = 1}$, ${n=29}$, and  ${\lambda = 12}$, (solid) and the shifted frequency response of the transfer function ${\hat{H}_{zw}}$ of the reduced order model obtained via balanced truncation for ${\nu = 3}$ (dashed), ${\nu = 4}$ (dashdotted), ${\nu = 5}$ (dotted), respectively. Note that assumption (A4) is satisfied by all reduced order models, as the magnitude of the shifted frequency response of the transfer function ${\hat{H}_{zw}}$ is below the lines defined  by ${\mu^{-1} - \epsilon_{\nu}}$. Figure~\ref{fig:bode_bt} only shows the (solid) line defined by ${\mu^{-1} = 0.1}$ as this is almost overlapped with the lines defined  by ${\mu^{-1} - \epsilon_{\nu}}$. Theorem~\ref{thm:1} guarantees that the original system is also strictly $2$-dominant, since balanced truncation preserves $2$-dominance of the linear dynamics and since ${\norm{H_{zw}}_{\infty,p}<\mu^{-1} - \epsilon_\nu }$ for ${\nu = 3, 4, 5}$. This is confirmed by the fact that the magnitude of the shifted frequency response of the transfer function ${{H}_{zw}}$ is below the (solid) line defined by ${\mu^{-1} = 0.1}$.

\section{Conclusion} \label{sec:conclusion}

A model reduction framework geared towards the analysis and design of systems that switch and oscillate has been described. Classical balanced truncation for linear, time-invariant systems has been revisited to develop a model reduction method for Lure systems which preserves dominance properties. The theory has been illustrated by the approximation of an oscillatory Lure system arising from the discretization of a heat flow control problem. 
A future research direction is to extend the proposed framework to systems that can be decomposed as the feedback interconnection of a linear system and a dominant nonlinear system using the small $\LTp$ gain theorem.

%\bibliography{refs}            

\end{document}